\documentclass[aip,reprint]{revtex4-2}
\usepackage{fdsymbol}
\usepackage{setspace}
\usepackage[utf8]{inputenc}
\usepackage[T1]{fontenc} 
\usepackage{xcolor,color,soul} 
\usepackage{graphicx}% Include figure files
\usepackage{bm}% bold math
\usepackage{color}
\usepackage{enumitem}
\usepackage[utf8]{inputenc}
\usepackage{wrapfig}
\usepackage{natbib}%[comma,sort&compress,square]
\usepackage[hidelinks, breaklinks=true, colorlinks=true, citecolor=blue, urlcolor=blue, linkcolor=blue]{hyperref}
\usepackage{framed}
\colorlet{shadecolor}{yellow}

\draft % marks overfull lines with a black rule on the right

\begin{document}

\title{Multiple Emulsions (W/O/W) for Confined Precipitation of Drug Nanoparticles}
\author{Umesh Dhumal}
\affiliation{Department of Chemical Engineering, Dr.~Babasaheb Ambedkar Technological University, Lonere, Raigad, Maharashtra, India 402103}
\email{umesh.dhumal007@gmail.com}
\date{\today}

\begin{abstract}
Multiple emulsions offer a route to confine nucleation and growth during drug precipitation, yet practical use is often limited by kinetic fragility and sensitivity to formulation and processing conditions. Here, we develop an ultrasound-assisted, two-stage emulsification strategy to generate water-in-oil-in-water (W/O/W) multiple emulsions with sufficient stability to enable formation of drug-rich submicron particulates. We first establish an operating window using simple W/O emulsions, showing that increased Tween~80 concentration and intensified sonication (higher amplitude and larger probe) yield smaller droplets and reduced coarsening tendencies. Using this window, W/O/W emulsions are formulated and systematically screened via surfactant pairing across ionic, non-ionic, and polymeric stabilizers. Ionic--non-ionic combinations provide the most favorable droplet-size control, with CTAB--Tween~80 emerging as a practically robust formulation. Cyclohexane is identified as a reproducible platform oil for downstream precipitation using the lead CTAB--Tween~80 formulation. Finally, curcumin-loaded W/O/W constructs generate curcumin-rich submicron particulates, supporting multiple emulsions as experimentally accessible microstructured environments for particle engineering of poorly soluble drugs.
\end{abstract}


\maketitle % must follow title, authors, abstract, and keywords (and \pacs if used)

\section*{Impact Statement}
Access to effective medicines is frequently limited by poor aqueous solubility of modern drug molecules, which can reduce and destabilize oral absorption and force higher doses, increasing cost and side effects. This work advances a practical, low-temperature route to improve drug dissolution performance by engineering drug-rich submicron particulates using water–oil–water (W/O/W) multiple emulsions as confined “microenvironments” for precipitation. We establish a two-stage, ultrasound-assisted emulsification workflow that decouples high-shear formation of a fine primary emulsion from low-shear transfer to the multiple-emulsion architecture, improving short-term kinetic robustness and preserving compartmentalization needed for controlled particle formation. Systematic screening identifies formulation and processing levers—surfactant concentration, ultrasound amplitude, probe geometry, and surfactant pairing—that jointly govern droplet stability and the ability to maintain confined domains. Using curcumin as a model poorly water-soluble compound, we demonstrate proof-of-concept formation of curcumin-rich submicron particulates within stabilized W/O/W constructs.

The approach is relevant to public health and sustainable manufacturing because it can reduce reliance on high-temperature or high-pressure particle-engineering routes, and it emphasizes scalable, experimentally accessible control variables (energy input, formulation composition). These features support translation toward cost-effective formulations that improve oral delivery of poorly soluble drugs, aligning with SDG 3 (Good Health and Well-Being) and SDG 9 (Industry, Innovation, and Infrastructure).

%\pacs{} % optional; leave empty or remove if not required by the journal

\section*{I. Introduction}
\label{sec:intro}

Poor aqueous solubility remains a common barrier to oral delivery for many pharmaceutically active compounds, because slow dissolution in gastrointestinal fluids can limit the rate and extent of absorption and contribute to variable exposure.\cite{kalepu_2015,lipinski_2001,savjani_2012}
From a transport perspective, dissolution kinetics depend on both the concentration driving force relative to saturation and the interfacial area available for mass transfer between the solid and surrounding fluid.\cite{noyes_1897}
Accordingly, enabling strategies for poorly soluble drugs pursue increased apparent solubility (e.g., salt/prodrug design and solubilizing excipients) and/or increased surface area via particle-size reduction.\cite{mueller_2009,liversidge_2008}
Particle-size reduction is particularly impactful for Biopharmaceutics Classification System (BCS) class~II compounds (low solubility, high permeability), where dissolution is frequently the controlling barrier to systemic exposure.\cite{amidon_1995,liversidge_2008}
Curcumin is a representative poorly water-soluble drug (from \emph{Curcuma longa}) that has attracted sustained interest due to reported antioxidant, anti-inflammatory, antimicrobial, and antitumor activities; however, its oral translation is strongly constrained by low and variable bioavailability, motivating extensive development of solubility- and dissolution-enhancement strategies including particle engineering and nano/micro-formulations.\cite{thorat_2014,anand_2007,hewlings_2017,mahran_2017}

Chemical approaches to address low aqueous solubility most commonly include salt formation and prodrug design.
Salt formation can markedly improve dissolution rate and apparent solubility for ionizable APIs and remains a mainstay of solid-form development; however, it is inherently limited to compounds with suitable acid--base functionality and can introduce developability risks such as salt disproportionation, hygroscopicity, and pH-dependent precipitation or solid-form instability during scale-up and storage.\cite{serajuddin_2007,gupta_2018,berge_1977}
Prodrug strategies can similarly improve biopharmaceutical performance by transiently masking problematic functionalities or tuning lipophilicity/permeability, but they require predictable bioconversion and can alter disposition; variability in activation, exposure, or metabolite profiles can add uncertainty to efficacy and safety assessment.\cite{mueller_2009,rautio_2008,huttunen_2011,rautio_2018}
Consequently, formulation-based particle engineering remains broadly applicable because it preserves molecular identity while improving dissolution performance; in this context, nanosuspensions and drug nanocrystals provide increased surface area (and often faster dissolution) without chemical modification, provided that stabilization against aggregation and growth is achieved.\cite{rabinow_2004,liversidge_2008,merisko_liversidge_2003,junghanns_2008,moeschwitzer_2013}

Top-down and bottom-up strategies are both used to access the sub-micron regime, but they impose different control challenges and failure modes.\cite{rabinow_2004,kipp_2004,junghanns_2008,moeschwitzer_2013}
Top-down methods (e.g., wet media milling and high-pressure homogenization) are attractive from a scalability and solvent-minimization standpoint and have become industrially mature routes for producing drug nanocrystals when appropriate stabilizers are present.\cite{keck_2006,peltonen_2010,junghanns_2008,moeschwitzer_2013}
However, the intense mechanical stresses and local temperature rise during comminution can induce surface disorder, partial amorphization, or polymorphic conversion, with consequences for redispersibility and physical stability.\cite{peltonen_2010,chikhalia_2006}
In contrast, bottom-up approaches (e.g., liquid antisolvent precipitation and related solvent-shift methods) create nanoparticles by generating high supersaturation and rapid nucleation; particle size and polydispersity are therefore governed by the competition between nucleation, growth, and aggregation on short time scales.\cite{thorat_2012,matteucci_2006,verma_2009,chen_2011}
In practice, these methods demand tight control of micromixing and supersaturation trajectories to avoid uncontrolled growth and agglomeration, and they may introduce solid-form and residual-solvent concerns if crystallization proceeds through metastable intermediates or solvent-mediated pathways.\cite{thorat_2012,matteucci_2006,verma_2009}
Across both routes, stabilizers (surfactants and/or polymers) are typically essential to provide steric/electrostatic barriers against aggregation and to mitigate growth phenomena such as Ostwald ripening; moreover, stabilizer identity and concentration can influence coarsening through solubilization effects and interfacial-layer properties.\cite{verma_2011,rabinow_2004,junghanns_2008}

Several conventional routes are used to produce drug micro-/submicron particulates, including thermal recrystallization, spray drying, and solvent--antisolvent precipitation.
In practice, these methods can struggle to consistently deliver small size together with narrow particle-size distributions and reproducible morphology at practical throughput.
A key reason is that particle formation is highly sensitive to local supersaturation histories and mixing. For drying-based routes, coupled evaporation and heat transfer further influence nucleation, growth, and final particle structure.\cite{kawabata_2011,thorat_2012}
In spray drying, for example, the final attributes depend on atomization and drying kinetics and may be limited by thermal exposure and collection inefficiencies as particle size decreases.\cite{vehring_2008,sosnik_2015}
More broadly, supercritical-fluid-based micronization routes can offer rapid solvent removal and tunable thermodynamics, but frequently rely on elevated pressures and specialized pressurization/expansion hardware (often including high-pressure pumps and engineered nozzles), which can complicate implementation and scale-up.\cite{jung_2001,martin_2008}

To address some of these constraints, Dalvi and Mukhopadhyay developed precipitation by pressure reduction of CO$_2$-expanded organic liquids (PPRGEL), where subcritical CO$_2$ expansion and rapid depressurization provide an additional handle to generate high transient supersaturation and drive particle formation without the same atomization pathway.\cite{dalvi_2009a,dalvi_2009b}
Subsequent modeling and analysis quantified the large transient cooling and rapid supersaturation generation during depressurization and suggested that interfacial phenomena (e.g., gas--liquid interface effects under vigorous CO$_2$ ebullition) can contribute under practical conditions.\cite{kumar_2013}
However, PPRGEL still requires high-pressure operation and CO$_2$ handling infrastructure and remains sensitive to depressurization and mixing/transport time scales, motivating complementary approaches that achieve controlled precipitation under milder processing conditions.
In this context, multiple emulsions offer a structurally defined alternative: their compartmentalized microstructure can act as a confined environment for solvent exchange and precipitation within droplet domains.\cite{jimenez_colmenero_2013,muschiolik_2007,garti_1997,vandergraaf_2005}

A complementary and conceptually distinct strategy is to structure the fluid environment in which supersaturation is generated.
Multiple emulsions (notably W/O/W) offer such an approach because their compartmentalized microstructure can function as a microstructured ``reaction environment'' for particle formation.\cite{jimenez_colmenero_2013,muschiolik_2007,garti_1997}
In these systems, solvent exchange and composition gradients across the oil ``membrane'' can be leveraged to drive nucleation and precipitation inside confined internal droplets.\cite{hou_1997,vandergraaf_2005}

A central challenge, however, is that multiple emulsions are intrinsically kinetically fragile.
Because they contain at least two interfaces, they can fail through internal/external coalescence, leakage, and mass-transfer-driven coarsening, including Ostwald ripening.\cite{florence_1981,iupac_2006,lifshitz_1961,kabalnov_2001}
For precipitation templating, stability is therefore not only a formulation objective; it is a prerequisite for reproducible particle formation because confinement must be maintained over the nucleation-and-growth time scale.\cite{vandergraaf_2005,garti_1997}

Stability is strongly formulation-dependent and commonly requires careful emulsifier pairing to rapidly protect newly created interfacial area at both the inner and outer interfaces.\cite{garti_1997,florence_1981}
Empirical blending concepts based on hydrophilic--lipophilic balance (HLB) have therefore been used to guide co-surfactant selection and improve multiple-emulsion robustness.\cite{shinoda_1980,hou_1997}
In addition, interfacial complexation strategies (e.g., macromolecule--surfactant association at the oil--water interface) have been explored to strengthen interfacial films and reduce leakage.\cite{law_1986}
Process choices can also improve control: membrane-based production methods have demonstrated narrower droplet-size distributions and improved uniformity for multiple emulsions, although equipment complexity and throughput remain practical constraints.\cite{vladisavljevic_2004,vladisavljevic_2006,dragosavac_2012,lindenstruth_2004}

High-intensity ultrasound provides a flexible process lever for controlling emulsion microstructure.
It generates fine droplets through cavitation-driven breakup, and droplet size can be tuned through energy input, processing time, and probe geometry.\cite{kaltsa_2013,li_2018}
Motivated by these considerations, we combine ultrasound-assisted formation of primary W/O emulsions with low-shear secondary emulsification to generate W/O/W multiple emulsions and evaluate their kinetic stability as precipitation templates for drug-rich submicron particulates.
The remainder of the paper is organized into Methods, Results and Discussion, and Conclusions and Future Directions.
Additional experimental details are provided in Section SI.

\section*{II. Methods}
\label{sec:methods}
\subsection*{II.1 Materials}
Curcumin was used as the model poorly water-soluble compound for nanoparticle precipitation studies. The oil phase for emulsion preparation employed toluene, carbon tetrachloride, or cyclohexane (depending on the experiment), and distilled water was used as the aqueous phase (inner and/or outer phase for multiple emulsions). Surfactants and polymeric stabilizers examined in different ionic/non-ionic/polymeric pairings included sodium dodecyl sulfate (SDS), cetyltrimethylammonium bromide (CTAB), polysorbate 80 (Tween~80), Poloxamer~407, hydroxypropyl methylcellulose (HPMC), and poly(vinyl pyrrolidone) (PVP). All chemicals were procured from S.D. Fine Chemicals Ltd.\ (India) and used as received unless otherwise specified.
\subsection*{II.2 Emulsion preparation and experimental design}
All experiments were conducted in batch mode using a two-stage emulsification strategy. In Stage~1, a primary water-in-oil (W/O) emulsion was prepared by probe sonication while the internal aqueous phase was added dropwise into the oil phase. In Stage~2, the primary W/O emulsion was dispersed into an external aqueous phase under low shear to form water-in-oil-in-water (W/O/W) multiple emulsions. During probe sonication, the vessel was maintained in an ice bath to limit bulk temperature rise.
The influence of key operating parameters was evaluated by varying (i) stabilizer concentration, (ii) sonication amplitude, and (iii) probe diameter in a controlled manner (details in Supplementary Methods). For multiple-emulsion stability screening, surfactants were selected to represent ionic, non-ionic, and polymeric stabilizers and were used in paired combinations across the oil and aqueous phases. Surfactant concentrations in each phase were set relative to the corresponding critical micelle concentration (CMC) using nominal levels of 10\%, 30\%, 50\%, 70\%, and approximately 100\% of CMC.

Droplet sizes were quantified by optical microscopy (Motic Pvt.\ Ltd.) coupled with image-based particle analysis (Biovis Image Analyser). For each sample, multiple micrographs were analyzed to obtain droplet-size distributions; droplet size was summarized using the minimum, maximum, mean, and standard deviation from image analysis. Short-term stability was monitored over 0--4~h with intermediate time points selected according to the experiment. Longer-term monitoring of selected formulations is provided in Supplementary Methods.

For precipitation experiments, curcumin (30~mg per batch) was incorporated into the oil phase prior to emulsification. W/O/W emulsions were then prepared using the same two-stage procedure. The resulting dispersions were monitored by optical microscopy and image-based analysis. Where solid recovery was required, precipitated solids were separated by centrifugation and dried by lyophilization.

\section*{III. Results and Discussion}
\label{sec:results}

The results are presented in three parts that mirror the experimental workflow.
First, we quantify how ultrasound processing parameters and stabilizer level control droplet size and short-term stability in simple W/O emulsions, and we use these data to define an operating window for preparing primary emulsions suitable for W/O/W transfer.
Second, we compare W/O/W stability across surfactant classes and pairings to identify formulations that minimize droplet growth and structural breakdown.
Third, we demonstrate curcumin precipitation within a stabilized W/O/W template as a proof-of-concept for emulsion-templated drug particle formation.

\subsection*{III.1 Simple W/O emulsions: Ultrasound parameter screening}
\label{subsec:wo_simple}

Simple water-in-oil (W/O) emulsions were used as a screening platform before preparing multiple emulsions.
This step had two objectives.
We first established a reproducible droplet-sizing workflow using optical microscopy and image analysis.
We then identified ultrasound conditions that yield small droplets with limited short-term growth, because the primary W/O morphology must survive the subsequent low-shear W/O/W formation step.

The model system comprised distilled water dispersed in carbon tetrachloride (CCl$_4$), stabilized with Tween~80.
Unless stated otherwise, emulsions were prepared by probe sonication for 10~min with the bulk temperature controlled at 30~$^\circ$C (ice-bath cooling during sonication).
Droplet sizes were quantified from microscopy using a BIOVIS workflow.
At least five micrographs were analyzed per time point to reduce sampling bias.
The screening focused on stabilizer concentration, ultrasound amplitude, and probe diameter.
These variables jointly determine (i) interfacial-area generation during breakup and (ii) how quickly the newly created interface becomes protected by surfactant adsorption.\cite{abismail_1999,behrend_2000,kaltsa_2013,li_2018}

\begin{figure}[t]
	\centering
	\includegraphics[width=\columnwidth]{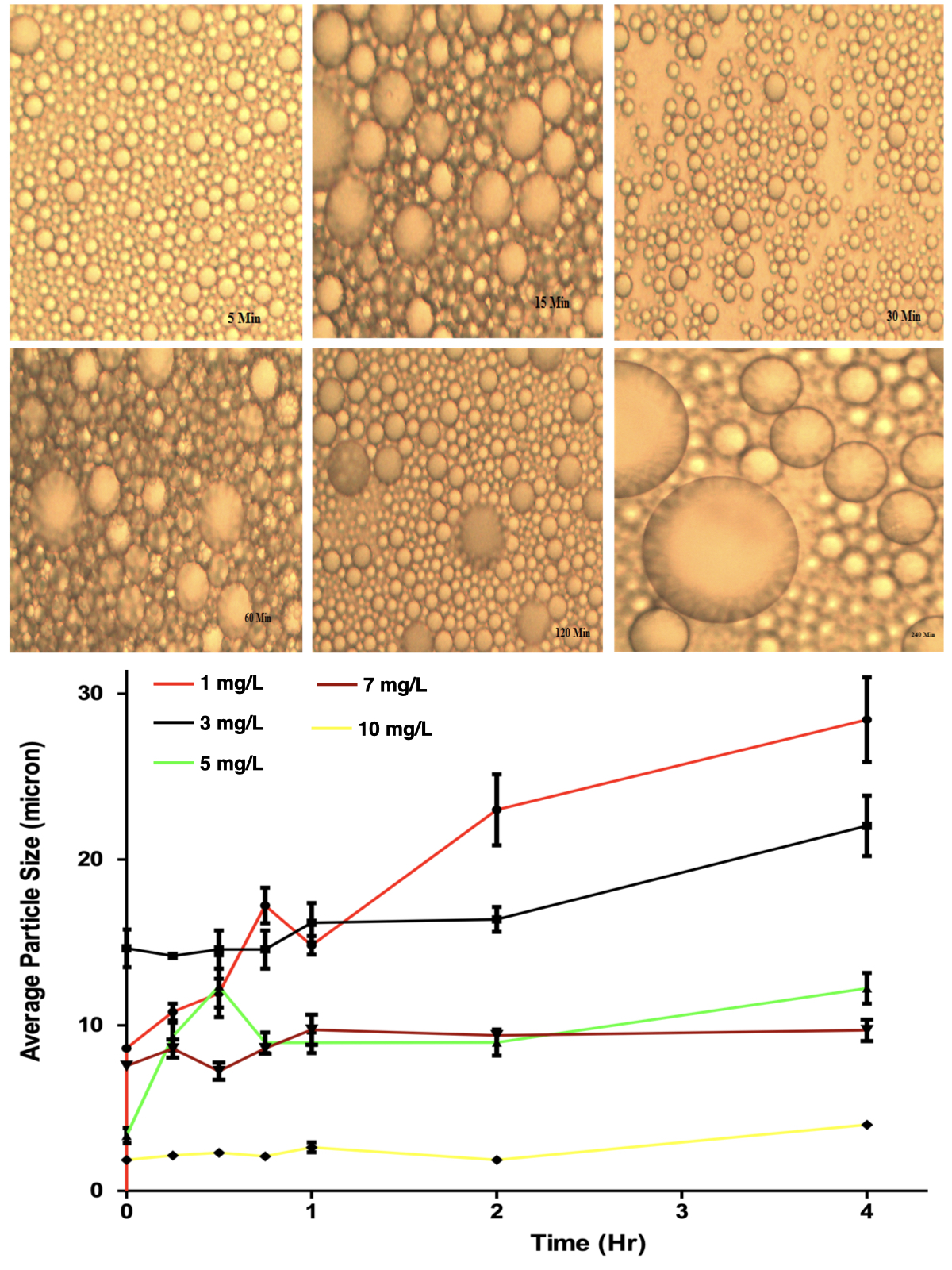}
	\caption{Effect of stabilizer concentration on W/O emulsion stability.
		Representative optical micrographs (top; selected aging times) and temporal evolution of the mean droplet size (bottom) for CCl$_4$/water W/O emulsions prepared by probe sonication (10~min, 20\% amplitude, 30~$^\circ$C) with Tween~80 concentrations of 1--10~mg/L.
		Error bars denote the dispersion obtained from image-based analysis of multiple micrographs per time point.}
	\label{fig:wo_stabconc}
\end{figure}

\subsubsection*{III.1.1 Effect of stabilizer concentration}

Tween~80 concentration was varied to quantify how interfacial coverage influences droplet size and early-time stability in W/O emulsions.
Concentrations of 1, 3, 5, 7, and 10~mg/L were tested.
All samples were prepared at fixed sonication time (10~min) and fixed amplitude (20\%).
Droplet sizes were monitored for several hours after preparation.

Figure~\ref{fig:wo_stabconc} shows a strong stabilizer dependence of both the initial droplet size and the subsequent size evolution.
At low Tween~80 levels (1--3~mg/L), the mean droplet size grows rapidly and reaches the tens-of-microns range over the 4~h window.
The micrographs in this regime show the progressive appearance of large droplets, indicating that a small population of coarsened/coalesced droplets can dominate the mean.
At higher Tween~80 levels (7--10~mg/L), droplets remain substantially smaller and the size evolution is weaker, indicating improved kinetic robustness.

The observed growth at low stabilizer loading is consistent with coupled destabilization processes.
Coalescence can occur when freshly created interfaces are not rapidly covered, especially immediately after sonication when collision rates are high.\cite{abismail_1999,behrend_2000}
In parallel, diffusional coarsening (Ostwald ripening) can increase the characteristic droplet size when Laplace-pressure differences drive molecular transport from smaller to larger droplets.\cite{iupac_2006,lifshitz_1961,kabalnov_2001,taylor_1998}
Ripening in emulsions is sensitive to the dispersed-phase solubility and to surfactant-mediated transport pathways.\cite{taylor_1998,weiss_2000,ariyaprakai_2010}
Increasing Tween~80 concentration plausibly improves stability by accelerating interfacial coverage during breakup and by producing a more persistent interfacial film that resists recoalescence.
Some residual evolution is still evident at intermediate concentrations (e.g., 5--7~mg/L), which motivates simultaneously increasing breakup efficiency (smaller initial droplets) and improving adsorption/coverage kinetics in the amplitude and probe studies that follow.\cite{verma_2011}

\subsubsection*{III.1.2 Effect of ultrasound amplitude (power)}

Ultrasound amplitude was varied to quantify how acoustic intensity controls droplet breakup and the resulting short-term stability.
The oil phase was CCl$_4$ and the dispersed phase was distilled water.
Tween~80 was fixed at 5~mg/L.
Emulsification was performed for 10~min at 30~$^\circ$C with amplitude settings of 20\%, 40\%, 60\%, 80\%, and 100\%.
Droplet sizes were quantified by microscopy immediately after emulsification and during aging.

\begin{figure}[t]
	\centering
	\includegraphics[width=\columnwidth]{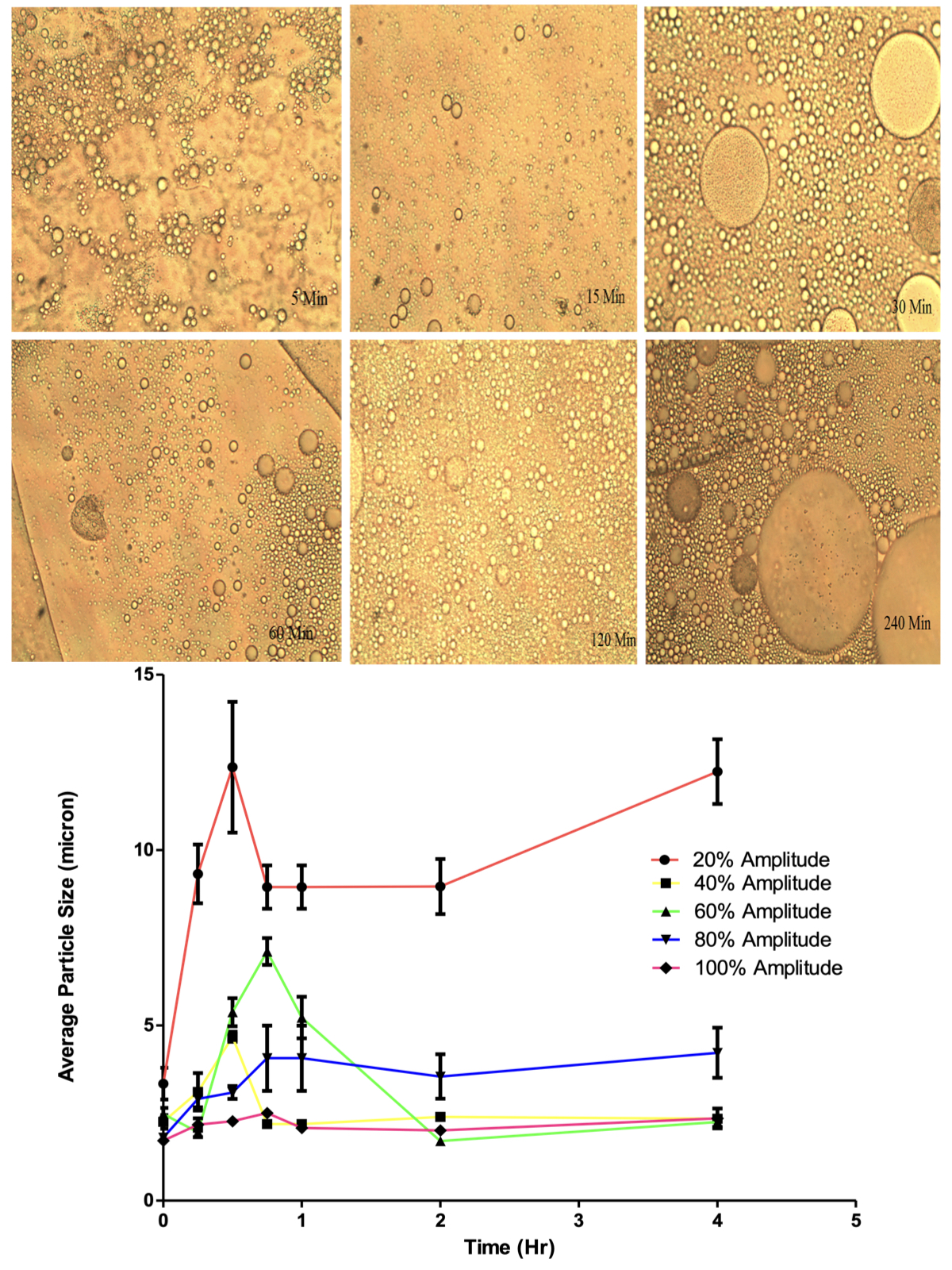}
	\caption{Effect of ultrasound amplitude on W/O emulsion evolution.
		Optical micrographs (top panels) illustrate droplet morphology at representative aging times, and the plot (bottom) shows mean droplet size as a function of time for emulsions prepared at different sonication amplitudes.
		Conditions: CCl$_4$ (oil) / water (dispersed), Tween~80 = 5~mg/L, sonication time = 10~min, temperature = 30~$^\circ$C.
		Error bars indicate the variability from image-based measurements.}
	\label{fig:wo_amplitude}
\end{figure}

Figure~\ref{fig:wo_amplitude} shows that increasing amplitude generally reduces the initial droplet size.
This trend is expected because higher acoustic intensity increases cavitation activity and enhances interfacial disruption near the probe tip.\cite{abismail_1999,behrend_2000,kaltsa_2013,li_2018}
The time evolution also depends strongly on amplitude.
At 20\% amplitude, the mean size increases sharply and remains high, consistent with weak breakup and poor kinetic robustness.
At high amplitudes (80--100\%), droplets remain in the low-micron regime with comparatively weak growth.

The intermediate-amplitude behavior is notably non-monotonic (especially the transient growth around $\sim$0.5--1~h for 60\%).
This pattern is consistent with a competition between droplet breakup and early-time recoalescence.
Breakup creates new interfacial area rapidly, but that interface must be stabilized on comparable timescales by surfactant adsorption.\cite{abismail_1999,behrend_2000,langevin_2014}
If adsorption lags behind area creation, transient recoalescence can generate a small number of larger droplets that disproportionately shift the mean.
Sampling effects can amplify this signature in microscopy-based means because a few large droplets carry high weight in number-averaged size metrics.
Practically, the data indicate that high-amplitude conditions provide a more reliable operating window under the present surfactant loading.
Accordingly, amplitudes in the 80--100\% range were carried forward for preparing primary W/O emulsions for W/O/W templating.

\subsubsection*{III.1.3 Effect of ultrasound probe diameter}

Probe diameter was varied because it changes the acoustic field, the cavitation zone, and the local energy density delivered to the fluid.
W/O emulsions were prepared under identical formulation and operating conditions using a small and a large probe.
Tween~80 was fixed at 5~mg/L.
Sonication was performed for 10~min at 20\% amplitude and 30~$^\circ$C.
Droplet sizes were monitored over time.

\begin{figure}[t]
	\centering
	\includegraphics[width=\columnwidth]{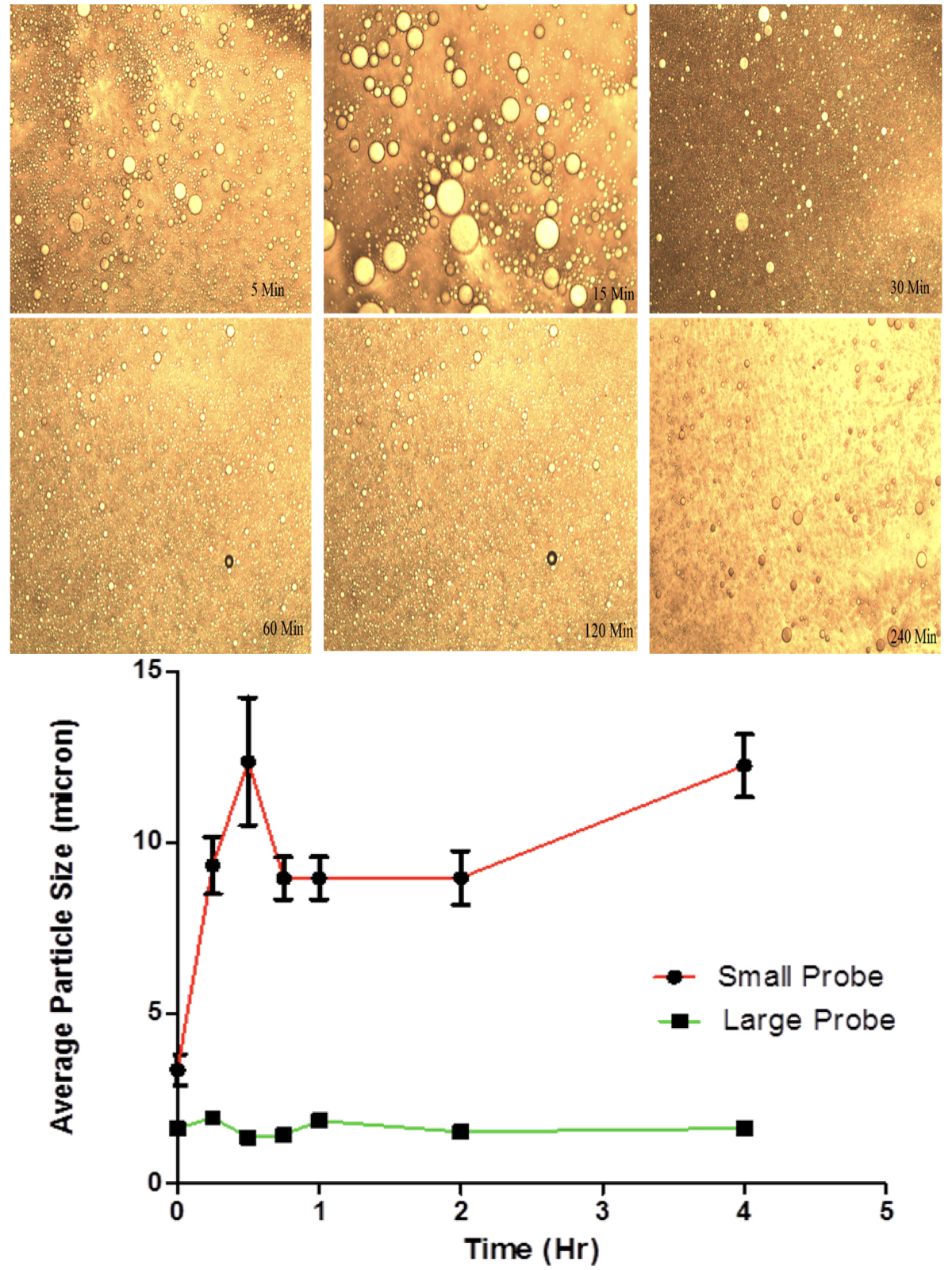}
	\caption{Effect of ultrasound probe diameter on W/O emulsion evolution.
		Representative optical micrographs (top panels) and temporal evolution of the mean droplet size (bottom) for CCl$_4$/water W/O emulsions prepared using a small vs.\ large sonication probe.
		Conditions: Tween~80 = 5~mg/L, sonication time = 10~min, amplitude = 20\%, temperature = 30~$^\circ$C.
		Error bars denote the variability obtained from image-based measurements of multiple micrographs per time point.}
	\label{fig:wo_probe}
\end{figure}

Figure~\ref{fig:wo_probe} shows that probe diameter has a first-order effect on droplet size.
The large probe produces droplets in the low-micron regime and maintains near-constant mean size over the observation window.
The small probe produces substantially larger droplets and exhibits strong growth, including the emergence of very large droplets at long times.
This behavior is consistent with established observations that ultrasonic emulsification depends not only on nominal amplitude settings but also on the spatial distribution of acoustic energy, which is controlled by probe geometry.\cite{abismail_1999,behrend_2000,kaltsa_2013,li_2018}

Overall, the W/O screening demonstrates that droplet size and short-term stability are jointly controlled by interfacial coverage (stabilizer availability and adsorption kinetics) and acoustic breakup intensity (amplitude and probe geometry).
These results define an operating window for preparing fine, robust primary emulsions prior to W/O/W formation.

\subsection*{III.2 W/O/W emulsions: stability screening by surfactant pairing}
\label{subsec:wow_screen}

W/O/W emulsions were prepared by a two-stage emulsification strategy.
A fine primary W/O emulsion was generated by probe sonication.
That primary emulsion was then dispersed into an external aqueous phase under low shear to preserve the internal droplet population.
This sequencing is standard for multiple emulsions because excessive shear in the second step can rupture the primary droplets and collapse the architecture.\cite{florence_1981,garti_1997,vandergraaf_2005}

Multiple emulsions contain at least two oil--water interfaces.
They are therefore vulnerable to coupled failure modes, including internal/external coalescence, leakage, and mass-transfer-driven coarsening.\cite{florence_1981,garti_1997,vandergraaf_2005}
Stability depends on surfactant adsorption kinetics, interfacial film strength, and (when polymers are used) the structure and rheology of mixed interfacial layers.\cite{langevin_2014}
Accordingly, we screened surfactant pairs spanning ionic, non-ionic, and polymeric stabilizers.
Formulations were ranked using time-resolved droplet-size evolution obtained by microscopy over 0--4~h.

\begin{figure}[t]
	\centering
	\includegraphics[width=\columnwidth]{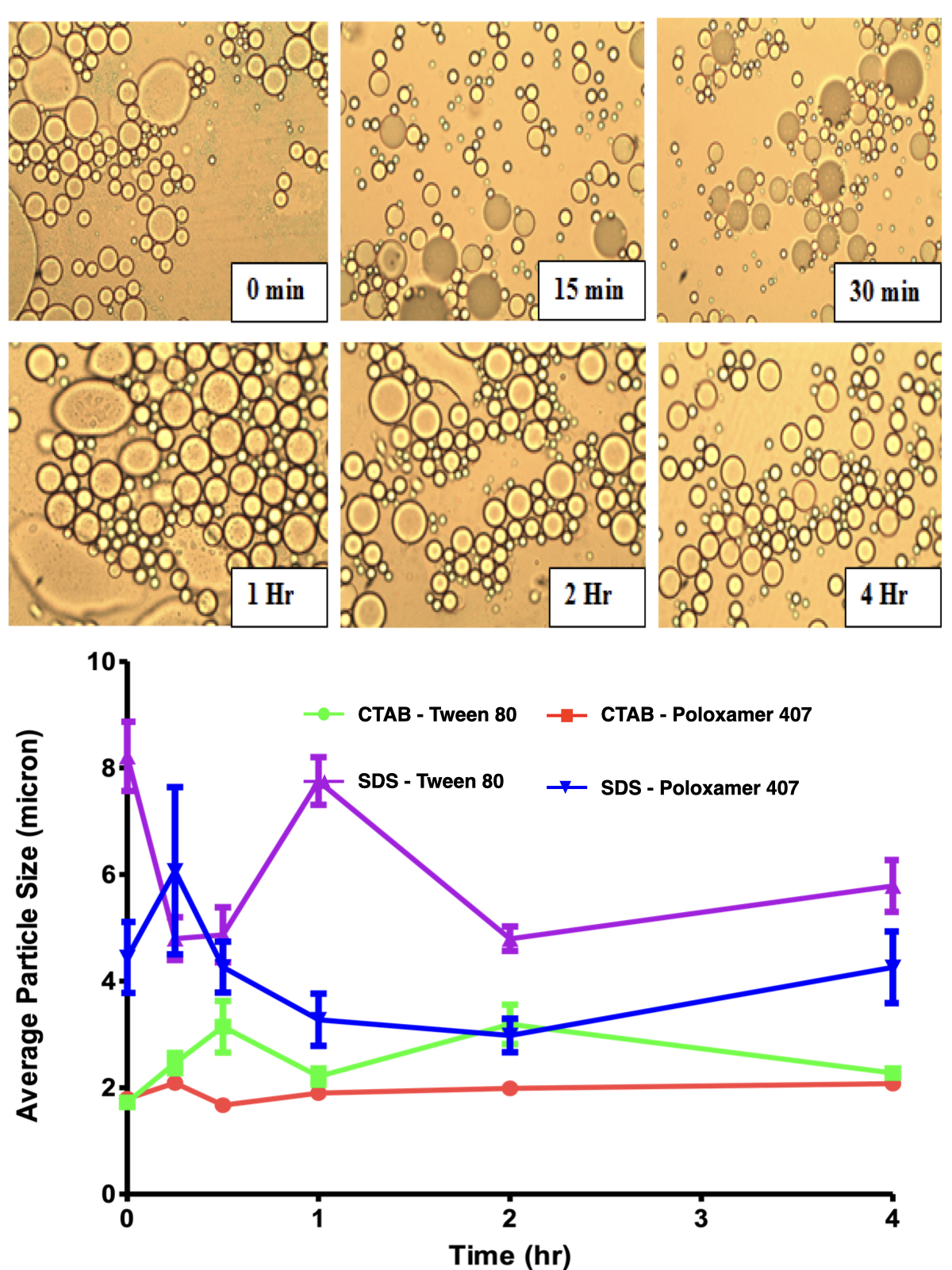}
	\caption{Ionic--non-ionic surfactant pairing in CCl$_4$-based W/O/W emulsions.
		Optical micrographs (0--4~h) and corresponding evolution of mean droplet size for CTAB--Tween~80, CTAB--Poloxamer~407, SDS--Tween~80, and SDS--Poloxamer~407 formulations.
		Error bars denote the standard deviation obtained from image-based analysis across multiple fields of view.}
	\label{fig:ionic_nonionic}
\end{figure}

\subsubsection*{III.2.1 Ionic--non-ionic surfactant pairing}

This set probes how ionic identity influences short-term stability when paired with non-ionic surfactants.
CTAB (cationic) and SDS (anionic) were evaluated in combination with Tween~80 and Poloxamer~407.
Figure~\ref{fig:ionic_nonionic} shows that CTAB-based pairings generate smaller droplets and maintain a comparatively steady mean size.
SDS-based systems show larger characteristic sizes and stronger time dependence, including visible structural coarsening in the micrographs.

A practical interpretation is that the CTAB-containing interfacial films provide more effective resistance to coalescence and leakage under the present conditions.
In multiple emulsions, this resistance must be achieved at both interfaces and must persist during early aging when transport and collisions continue.\cite{florence_1981,garti_1997,vandergraaf_2005}
Differences in adsorption kinetics and in interfacial-layer rheology can produce large differences in early-time robustness, even when the nominal droplet sizes immediately after preparation are similar.\cite{langevin_2014}
Among the CTAB-based systems, CTAB--Tween~80 provided consistently small droplets and robust microstructural integrity across the window.
It was therefore selected as the lead ionic--non-ionic formulation for subsequent tests.

\begin{figure}[t]
	\centering
	\includegraphics[width=\columnwidth]{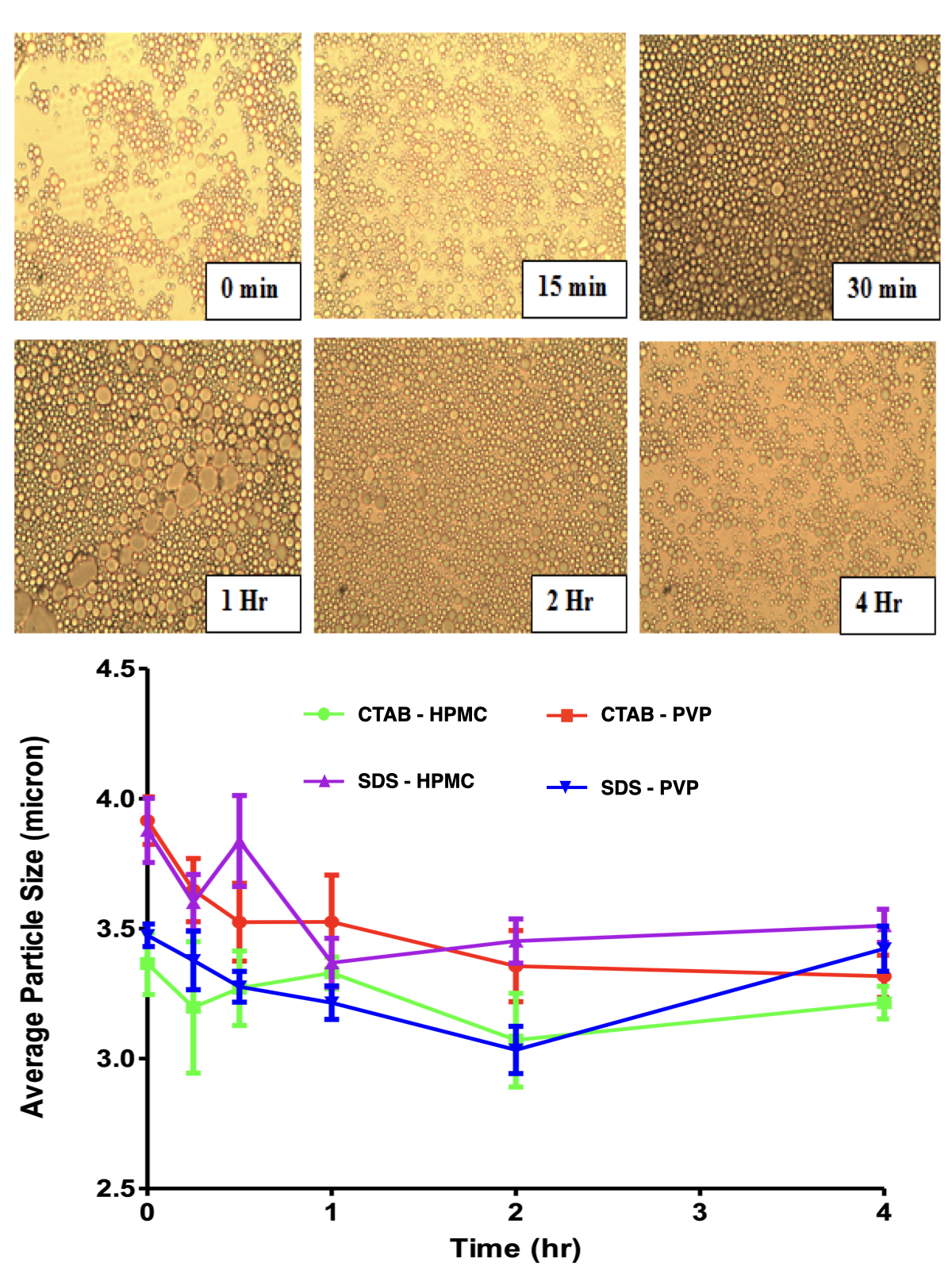}
	\caption{Ionic--polymeric surfactant screening for W/O/W multiple emulsions.
		Representative optical micrographs (0--4~h) and corresponding evolution of mean droplet size for CTAB--HPMC, CTAB--PVP, SDS--HPMC, and SDS--PVP systems.
		Error bars indicate the variability obtained from replicate image analysis.}
	\label{fig:ionic_polymeric}
\end{figure}

\subsubsection*{III.2.2 Ionic--polymeric surfactant pairing}

This set tests whether adding a polymeric stabilizer in the outer aqueous phase improves robustness by steric stabilization and modified interfacial rheology.
Figure~\ref{fig:ionic_polymeric} shows that all four formulations remain in a relatively narrow size band ($\sim$3--4~$\mu$m) with modest evolution.
CTAB--PVP yields the smallest mean size and the weakest drift.
SDS-based pairings show slightly larger means and greater variability.

Polymeric additives can stabilize multiple emulsions through several coupled mechanisms.
They can increase continuous-phase viscosity and reduce collision frequency.
They can also modify interfacial films through competitive adsorption or polymer--surfactant association, which changes interfacial elasticity and resistance to rupture.\cite{langevin_2014}
Within the tested design space, CTAB--PVP was the most robust ionic--polymeric candidate.

\begin{figure}[t]
	\centering
	\includegraphics[width=\columnwidth]{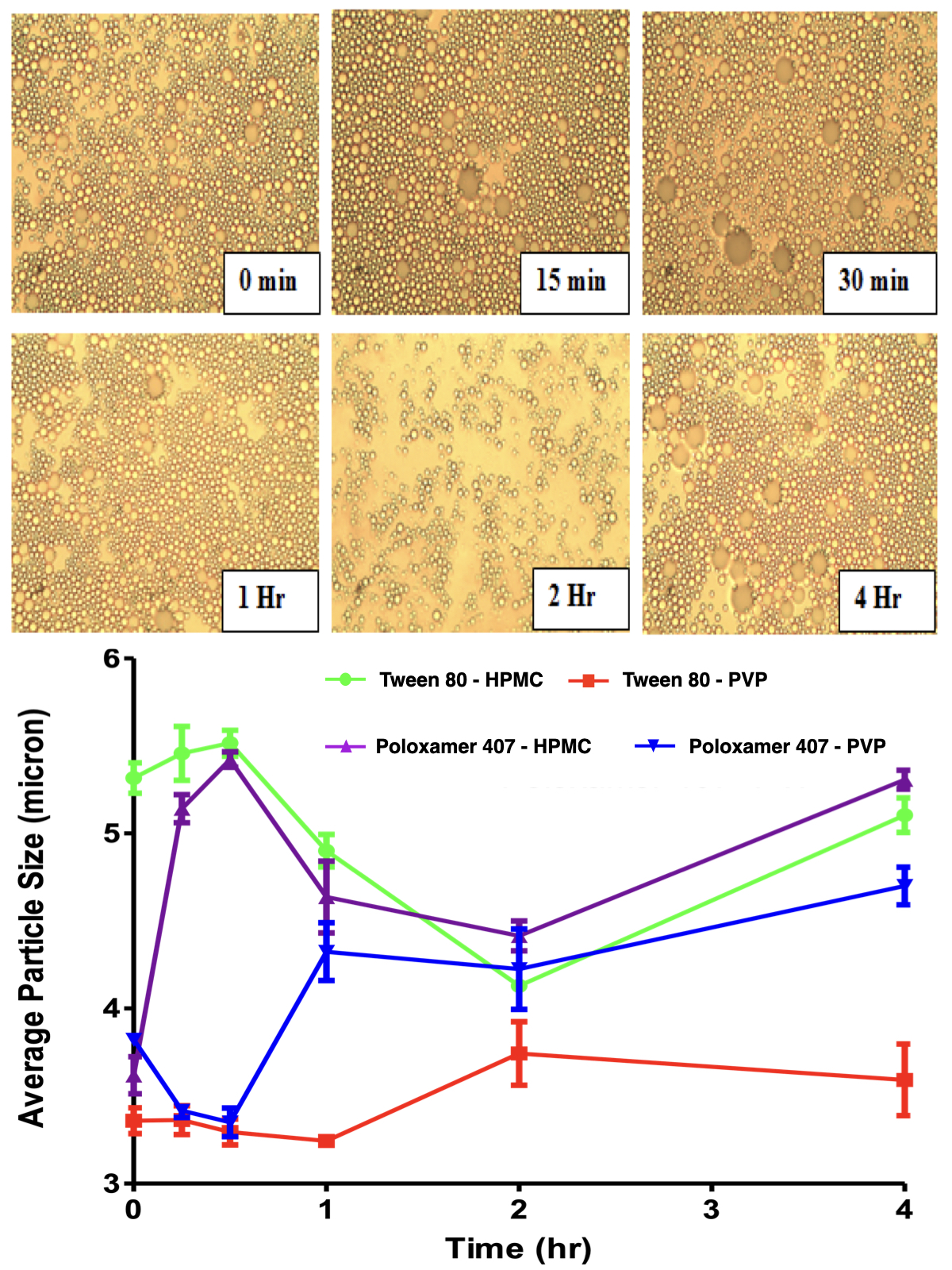}
	\caption{Non-ionic--polymeric surfactant screening for W/O/W emulsions.
		Representative optical micrographs (0--4~h) and evolution of mean droplet size for Tween~80--HPMC, Tween~80--PVP, Poloxamer~407--HPMC, and Poloxamer~407--PVP.
		Error bars indicate the variability obtained from replicate image analysis.}
	\label{fig:nonionic_polymeric}
\end{figure}

\subsubsection*{III.2.3 Non-ionic--polymeric surfactant pairing}

This set evaluates whether purely steric stabilization can provide robust W/O/W constructs.
Figure~\ref{fig:nonionic_polymeric} shows that Tween~80--PVP produces the smallest droplets and the most uniform microstructure.
Formulations containing HPMC are larger and exhibit clearer early-time growth.
Poloxamer-based systems also show a transient increase at early times.

A plausible explanation is kinetic.
During secondary emulsification, rapid adsorption reduces the time window in which partially covered droplets can coalesce.\cite{florence_1981,garti_1997}
Differences in adsorption kinetics and interfacial-layer rheology between Tween~80 and Poloxamer~407, and between PVP and HPMC, can therefore translate into measurable differences in early-time droplet growth.\cite{langevin_2014}
Across the full screening matrix, the results highlight two recurring requirements.
Interfacial coverage must form quickly and persist at both interfaces.
Droplets must also experience an effective barrier to droplet--droplet interactions during aging.
CTAB--Tween~80 was retained as the lead formulation because it provides both small droplet sizes and robust short-term stability.

\subsubsection*{III.3 Oil-phase sensitivity: cyclohexane-based W/O/W emulsions with CTAB--Tween~80}
\label{subsubsec:cyclohexane_ctab_tween}

After identifying CTAB--Tween~80 as the most robust pairing in the screening step, we evaluated oil-phase sensitivity by transferring the W/O/W protocol to cyclohexane as the membrane phase.
The surfactant concentration was varied systematically (1, 3, 5, 7, and 10~mg/L), with matched concentrations in the inner and outer aqueous phases.

\begin{figure}[t]
	\centering
	\includegraphics[width=\columnwidth]{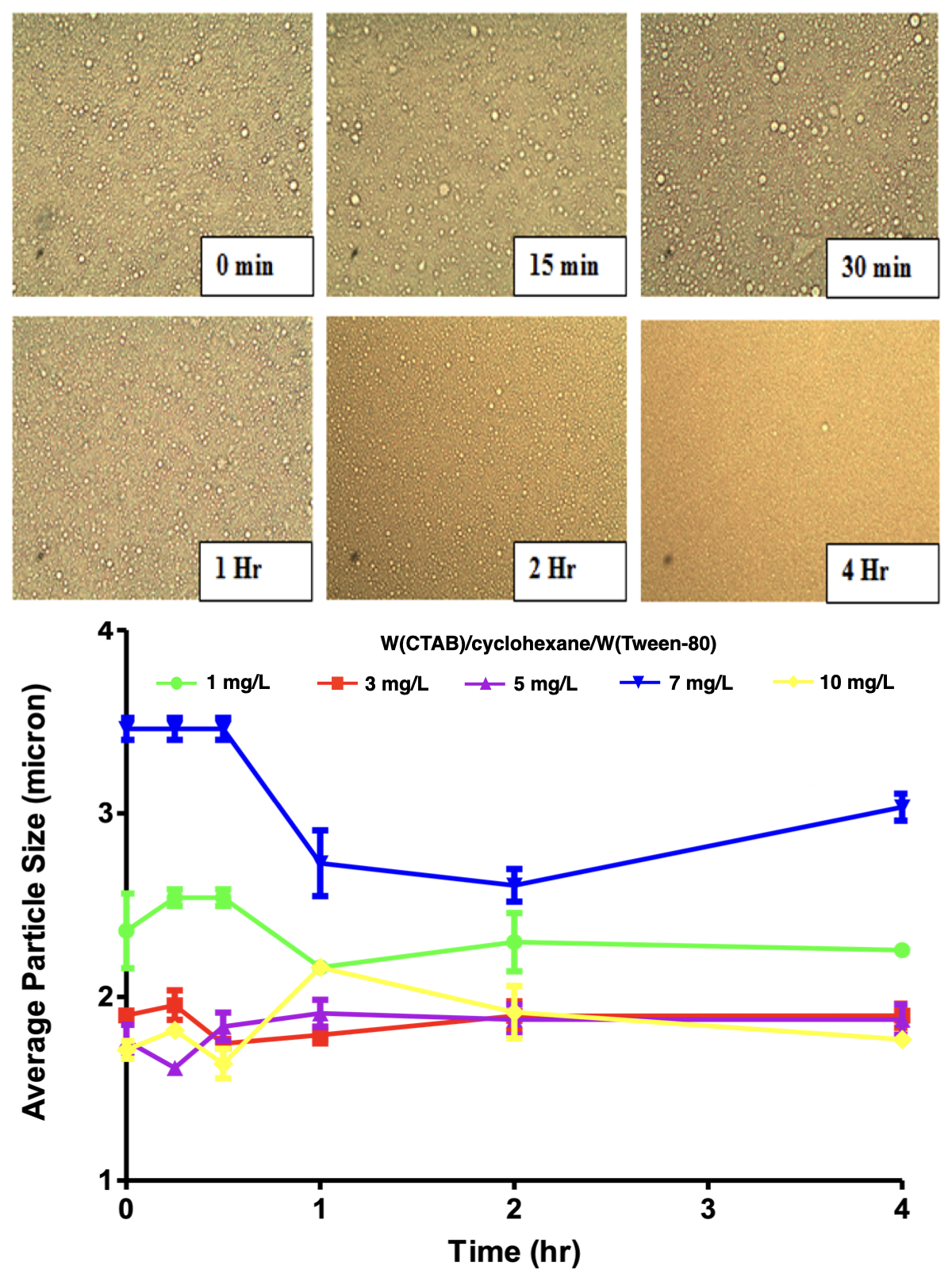}
	\caption{Oil-phase sensitivity of the CTAB--Tween~80 formulation in cyclohexane-based W/O/W emulsions (H$_2$O/cyclohexane/H$_2$O).
		Representative optical micrographs (selected times) and the corresponding evolution of mean droplet size are shown for matched surfactant concentrations (1--10~mg/L) in the inner and outer aqueous phases.}
	\label{fig:fig7}
\end{figure}

Figure~\ref{fig:fig7} shows that cyclohexane supports stable W/O/W constructs in the low-micron regime across the tested concentration window.
The lowest concentration (1~mg/L) yields larger droplets and stronger early-time growth, consistent with incomplete interfacial coverage.
Higher concentrations (3--10~mg/L) generally yield smaller droplets and weaker drift, although the concentration dependence is not strictly monotonic (e.g., the 7~mg/L condition is an outlier with a larger mean).
Such non-monotonicity is plausible in multi-interface systems because stability depends on how surfactant partitions between bulk, the two interfaces, and any aggregates, which can shift interfacial tension and film rheology in nontrivial ways.\cite{langevin_2014}
More broadly, oil choice can influence coarsening because ripening rates depend on oil solubility and the associated molecular-transport pathways.\cite{taylor_1998,weiss_2000}

\subsection*{III.4 Curcumin precipitation within W/O/W multiple-emulsion templates}
\label{subsec:curcumin_precip}

After establishing a robust W/O/W formulation window, the multiple-emulsion platform was used as a confined environment for precipitation of drug-rich particulates using curcumin as a model poorly water-soluble compound.
Curcumin was incorporated into the oil phase prior to emulsification, and W/O/W emulsions were prepared using the same two-stage protocol.
The working concept is that the compartmentalized W/O/W microstructure localizes composition gradients and solvent exchange across the oil layer, enabling supersaturation and nucleation within confined domains rather than in the fully mixed bulk.\cite{garti_1997,vandergraaf_2005,thorat_2012}

\begin{figure}[t]
	\centering
	\includegraphics[width=\columnwidth]{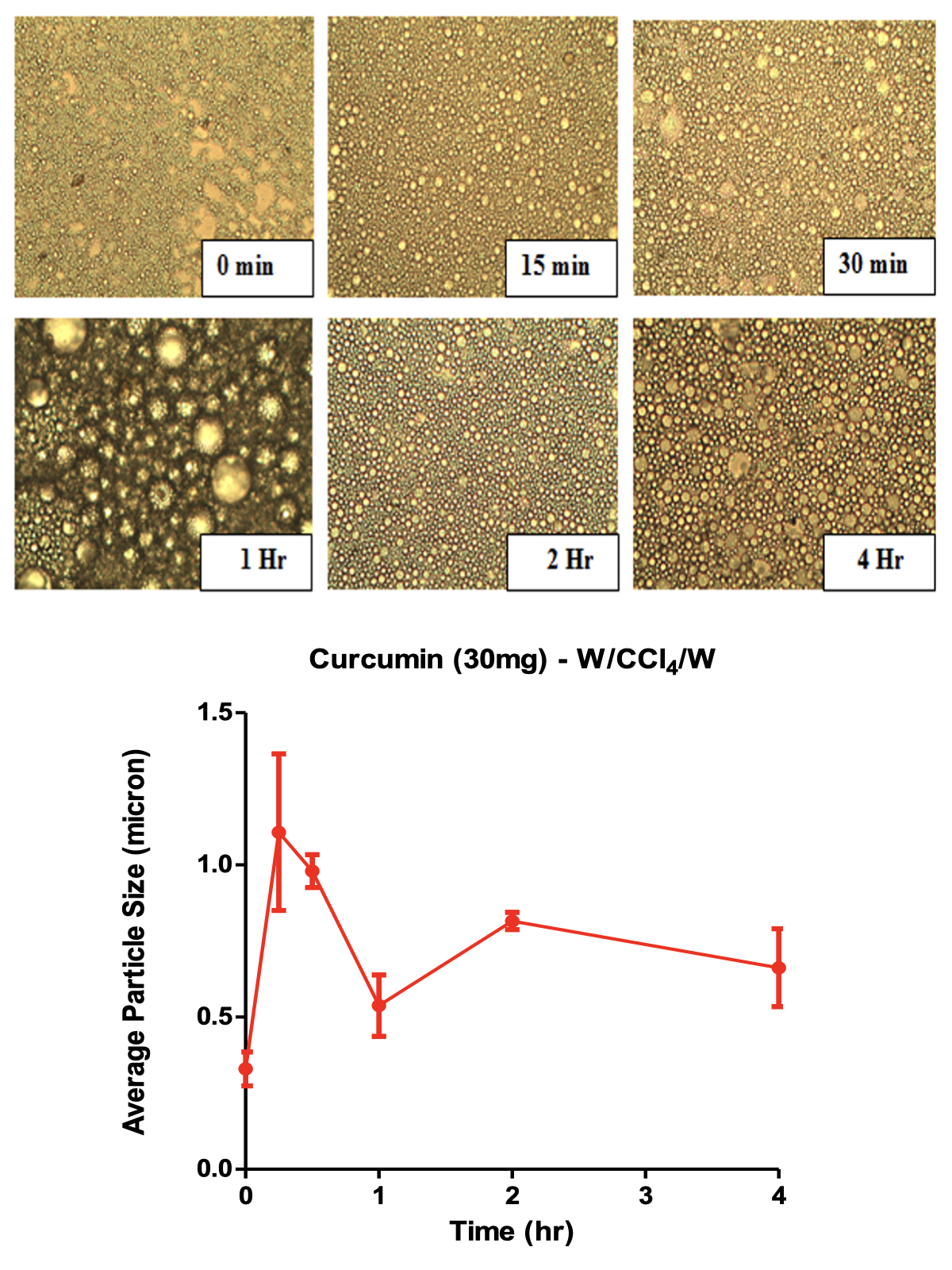}
	\caption{Curcumin precipitation within W/O/W multiple-emulsion templates.
		Representative optical micrographs (selected times) and evolution of mean particulate size following W/O/W formation with curcumin incorporated in the oil phase prior to emulsification.}
	\label{fig:curcumin_precip}
\end{figure}

Figure~\ref{fig:curcumin_precip} shows the emergence of curcumin-rich particulates in the submicron-to-micron range.
The apparent mean particulate size exhibits a non-monotonic evolution, with an early increase followed by a decrease and partial stabilization.
This behavior is consistent with multiple coupled processes acting during early aging.
A nucleation burst and early growth can increase the apparent mean size.
Subsequent restructuring can decrease it, for example through redistribution of curcumin between phases, breakup of weak aggregates during sampling, or changes in optical contrast as the internal composition evolves.
Coarsening can also compete, particularly if solubilization/transport pathways remain active.\cite{thorat_2014,thorat_2012,taylor_1998,weiss_2000}

While optical microscopy cannot fully resolve the smallest nanoscale population, these observations establish the key proof-of-concept.
Stable W/O/W constructs can serve as precipitation templates that yield curcumin-rich submicron particulates under comparatively mild processing conditions.
The approach is complementary to liquid antisolvent precipitation, which can produce curcumin nanoparticles but is highly sensitive to micromixing histories and stabilization dynamics.\cite{thorat_2014,thorat_2012,matteucci_2006}
Taken together, the results motivate multiple emulsions as a tunable microstructured environment in which formulation control (surfactant pairing and concentration) and process control (ultrasound-defined primary droplet size) jointly determine both the attainable size range and the short-term robustness required for reproducible particle formation.

\section*{IV. Conclusions and future directions}
\label{sec:conclusion_future}

This study demonstrates an ultrasound-assisted, two-stage emulsification strategy for generating W/O/W multiple emulsions with sufficient short-term kinetic stability to act as practical templates for precipitation of drug-rich submicron particulates.\cite{garti_1997,vandergraaf_2005}
The workflow intentionally decouples (i) high-shear formation of the primary W/O microstructure from (ii) low-shear transfer into a W/O/W architecture.
This separation is important because secondary high shear can rupture primary droplets, accelerate leakage, and erase compartmentalization before precipitation is complete.\cite{florence_1981,garti_1997,vandergraaf_2005}

The W/O screening results show that droplet size and early-time stability are controlled jointly by breakup intensity and interfacial protection.
Increasing ultrasound amplitude and using a larger probe reduce initial droplet size by increasing cavitation activity and local hydrodynamic stresses near the tip.\cite{suslick_1990,abismail_1999,behrend_2000,kaltsa_2013,li_2018}
However, breakup alone is not sufficient.
Ultrasound continuously generates new interfacial area, and that area must be rapidly covered by surfactant to prevent recoalescence of newly formed daughter droplets.\cite{abismail_1999,behrend_2000}
Accordingly, increasing Tween~80 concentration decreased the mean droplet size and reduced the magnitude of time-dependent growth.
This is consistent with faster and more complete interfacial coverage, together with reduced susceptibility to coupled destabilization mechanisms (collision-driven coalescence and diffusion-driven coarsening).\cite{iupac_2006,lifshitz_1961,kabalnov_2001}

For W/O/W systems, surfactant pairing across the two interfaces is the dominant determinant of kinetic robustness.\cite{garti_1997,florence_1981,vandergraaf_2005}
Multiple emulsions can fail through several coupled pathways.
These include (i) rupture of the oil film and leakage of internal droplets, (ii) internal or external coalescence, and (iii) droplet growth driven by compositional gradients, Laplace-pressure differences, and solute transport.\cite{garti_1997,opawale_1998,jiao_2002}
These pathways are sensitive to interfacial film strength and interfacial rheology, which depend on surfactant identity, mixed adsorption, and (when present) polymer--surfactant interactions.\cite{opawale_1998,jiao_2002}
Within the screening matrix explored here, the ionic--non-ionic family provided the best overall droplet-size control, and CTAB--Tween~80 emerged as a practically robust pairing over the 0--4~h window.
The oil-phase sensitivity tests further indicate that cyclohexane can provide reproducible W/O/W morphology when surfactant concentrations are sufficiently high to stabilize both interfaces and suppress early-time restructuring.\cite{garti_1997,florence_1981,vandergraaf_2005}

The curcumin precipitation experiments provide proof-of-concept that stable W/O/W constructs can function as experimentally accessible microstructured environments for particle engineering.
In this templating picture, precipitation is driven by solvent/solute redistribution across the oil ``membrane,'' which creates localized supersaturation within confined domains.
Confinement can reduce the spatial scale of mixing and can limit uncontrolled aggregation by restricting where nucleation and growth occur.\cite{vandergraaf_2005,thorat_2012,thorat_2014}
The observed time-dependent size evolution is consistent with competing processes during early aging, including ongoing mass transfer, nucleation/growth, and coarsening phenomena such as Ostwald ripening that are known to operate in disperse systems.\cite{iupac_2006,lifshitz_1961,kabalnov_2001}
Taken together, the results support a practical route to couple ultrasound-defined primary microstructure with formulation-guided stability control to enable multiple-emulsion templating for drug particle precipitation under comparatively mild conditions.

Future work should extend characterization beyond optical microscopy to more fully resolve the smallest size fraction and to separate droplet-scale evolution from solid particulate evolution.
DLS/NTA and electron microscopy would provide complementary resolution of the nanoscale population and particle morphology.
Additional experiments should quantify longer-time stability and decouple leakage/coalescence from diffusion-driven coarsening by independently measuring internal-phase retention and droplet-size distributions.\cite{garti_1997,opawale_1998}
Finally, generality should be tested across additional poorly water-soluble drugs and pharmaceutically acceptable solvent/surfactant constraints, with particular attention to scalability of ultrasound processing and to energy-density-based transferability of operating windows.\cite{modarres_gheisari_2019,liversidge_2008}

Detailed experimental protocols and additional supporting data are provided in the Supplementary Material.

\section*{Acknowledgments}
We would like to acknowledge the support of Technical Education Quality Improvement Programme (TEQUIP) for the fellowship to undertake this work. 
	
\section*{Declaration of Competing Interests}
The author declare no known competing financial interests or personal relationships that could have appeared to influence the work reported in this paper.

\section*{Data Availability Statement}
That data that support the findings of this study are available from the corresponding author upon reasonable request.

\section*{SI. Supplementary Methods}
\label{sec:supp_methods}
	
	\subsection*{Preparation of simple W/O emulsions for ultrasound parameter screening}
	Simple W/O emulsions were prepared to quantify the influence of ultrasound operating parameters on droplet size and short-term stability. Carbon tetrachloride was used as the oil phase and distilled water as the dispersed phase. Tween~80 served as the stabilizer in these screening studies.
	
	In a typical run, 50~mL of carbon tetrachloride containing Tween~80 at the desired concentration was placed in a beaker. A separate aqueous phase was prepared with distilled water containing Tween~80 at the same concentration. Prior to emulsification, both phases were equilibrated in a temperature-controlled bath at 30~$^\circ$C for 10~min. The oil phase was then subjected to probe sonication (Sonics), while the aqueous phase was introduced dropwise using a micropipette during sonication. Sonication was continued for a total of 10~min. An ice bath surrounding the vessel was used to limit bulk temperature rise during probe operation. The amplitude setting was selected according to the experimental design below, and the instrument-reported energy input was recorded for each run. Samples were collected immediately after emulsification and at subsequent time points for droplet-size analysis.
	
	\subsection*{Ultrasound parameter study design}
	Three parameters were varied systematically:
	\begin{enumerate}
		\item \textbf{Tween~80 concentration:} 1, 3, 5, 7, and 10~mg/L at fixed amplitude of 20\% and sonication time of 10~min.
		\item \textbf{Sonication amplitude:} 20, 40, 60, 80, and 100\% at fixed Tween~80 concentration of 5~mg/L and sonication time of 10~min.
		\item \textbf{Probe diameter:} two probe diameters (``small'' and ``large'') at fixed Tween~80 concentration of 5~mg/L, amplitude of 20\%, and sonication time of 10~min.
	\end{enumerate}
	
	\subsection*{Preparation of W/O/W multiple emulsions}
	Multiple emulsions were prepared using the two-stage framework. Depending on the experimental set, the oil phase was selected from toluene, carbon tetrachloride, or cyclohexane. The oil phase contained the designated lipophilic (oil-soluble) surfactant, while the external aqueous phase contained the designated hydrophilic (water-soluble) surfactant; distilled water was used for both the internal and external aqueous phases.
	
	\paragraph*{Stage 1: Primary W/O emulsion.}
	In a typical preparation, 50~mL of oil phase containing the lipophilic surfactant was placed under the ultrasonic probe and sonicated while 50~mL of the internal aqueous phase (distilled water) was added dropwise by micropipette. Unless stated otherwise, the primary-emulsion step for multiple-emulsion preparation was conducted for 10~min at 100\% amplitude. An ice bath was used during probe sonication to control bulk temperature. The instrument-reported energy input was recorded.
	
	\paragraph*{Stage 2: Secondary emulsification (W/O/W formation).}
	The external aqueous phase (50~mL) containing the hydrophilic surfactant was stirred using a magnetic stirrer. The freshly prepared W/O emulsion was then added dropwise to the external aqueous phase over 10~min under low shear to form the W/O/W multiple emulsion while minimizing rupture of the primary droplets.
	
	\subsection*{Surfactant-pair screening strategy}
	Surfactants were selected to represent ionic (SDS, CTAB), non-ionic (Tween~80, Poloxamer~407), and polymeric (PVP, HPMC) stabilizers and were used in paired combinations across the oil and aqueous phases. For multiple-emulsion stability screening, surfactant concentrations in each phase were set relative to the corresponding critical micelle concentration (CMC), using nominal levels of 10\%, 30\%, 50\%, 70\%, and approximately 100\% of CMC.
	
	\subsection*{Droplet-size analysis and stability protocol}
	Droplet sizes were quantified by optical microscopy (Motic Pvt.\ Ltd.) coupled with image-based particle analysis (Biovis Image Analyser). For each sample, multiple micrographs were analyzed to obtain droplet-size distributions; droplet size was summarized using the minimum, maximum, mean, and standard deviation returned by image analysis. Short-term stability measurements were performed over 0--4~h with intermediate time points selected according to the experiment. Longer-term monitoring of selected formulations is provided in Supplementary Methods.
	
	\subsection*{Drug precipitation and solid recovery}
	Curcumin was used as a model poorly water-soluble drug. For precipitation experiments, curcumin (30~mg per batch) was incorporated into the oil phase prior to emulsification. W/O/W emulsions were then prepared using the two-stage procedure described above, and the resulting dispersions were monitored by optical microscopy and image-based analysis. Where solid recovery was required, precipitated solids were separated by centrifugation and dried by lyophilization.
	
	% Reset figure numbering in the Appendix
	\setcounter{figure}{0}
	\renewcommand{\thefigure}{A\arabic{figure}}

	\clearpage
	
\bibliographystyle{aipnum4-2}
%\bibliography{references} 

%aipnum4-2.bst 2019-01-14 (MD) hand-edited version of apsrev4-1.bst

\end{document}